
\magnification=1200
\hsize=14.4truecm
\vsize=23.1truecm
\baselineskip=21pt plus 1pt minus 1pt
\parindent=1truecm
\centerline {\bf Shape Analysis of the Level Spacing Distribution}
\centerline {\bf around the Metal Insulator Transition}
\centerline {\bf in the Three Dimensional Anderson Model}
\bigskip
\bigskip
\centerline {Imre Varga$^1$, Etienne Hofstetter$^{2*}$, Michael
    Schreiber$^3$, and J\'anos Pipek$^1$}
\medskip\it
\centerline {$^1$Quantum Theory Group, Institute of Physics, Technical
    University of Budapest,}
\centerline {H--1521 Budapest, Hungary}
\centerline {$^2$Institut f\"ur Physikalische Chemie,
    Johannes-Gutenberg-Universit\"at,}
\centerline {Jakob-Welder-Weg 11, D-55099 Mainz, Germany}
\centerline {$^3$Institut f\"ur Physik, Technische Universit\"at
    Chemnitz--Zwickau,}
\centerline {Postfach 964, D-09009 Chemnitz, Germany}
\rm
\bigskip
\bigskip
    {\bf Abstract} - We present a new method for the numerical
    treatment of second order phase transitions using the level
    spacing distribution function $P(s)$. We show that the quantities
    introduced originally for the shape analysis of eigenvectors can
    be properly applied for the description of the eigenvalues as
    well. The position of the metal--insulator transition (MIT) of the
    three dimensional Anderson model and the critical exponent are
    evaluated. The shape analysis of $P(s)$ obtained numerically shows
    that near the MIT $P(s)$ is clearly different from both the
    Brody distribution and from Izrailev's formula, and the best
    description is of the form $P(s)=c_1\,s\exp(-c_2\,s^{1+\beta})$,
    with $\beta\approx 0.2$. This is in good agreement with recent
    analytical results.

\vfil
\eject
\centerline {\bf I. INTRODUCTION}
\medskip

    Recently a novel method has been introduced for the location of
    the critical point and the determination of the critical exponent
    in the three dimensional (3D) Anderson model exhibiting a
    metal-insulator transition (MIT). It has been demonstrated by
    Shklovskii {\it et al.}$^1$ and by Hofstetter and
Schreiber$^{2,3}$
    that random matrix theory (RMT)$^4$ may serve as a tool for a
    surprisingly accurate calculation. This time the necessary
    information is derived using the spectrum rather than the wave
    functions of the system. It is also expected that this new method
    is easily applicable for other types of second order phase
    transitions.

    The novel approach is based on the study of the statistical
    properties of the eigenvalues of the Hamiltonian on both sides of
    the MIT. Based on an analogy between the kicked rotator and the
    Anderson model$^5$ the MIT can be considered as a transition from
    the chaotic regime to the non-chaotic one, or in other words using
    the terminology of the RMT, from the Gaussian orthogonal ensemble
    (GOE) to the Poisson ensemble (PE) of random matrices. The level
    spacing distributions for both ensembles are known: for the GOE it
    is very well described by the Wigner surmise
    $$P_{\rm GOE}(s)\cong {\pi\over 2}\,s\,
                     \exp\left (-{\pi\over 4}s^2\right ),   \eqno(1)$$
    that shows linear level repulsion for low $s$. For the PE one has
    $$P_{\rm PE}(s)=\exp (-s),                              \eqno(2)$$
    i.e. in this case the energy levels of localized states may be
    arbitrarily close.

    The model under consideration is described by the usual
    tight--binding Hamiltonian
    $${\cal H}=\sum_i\varepsilon_i |i><i|
              +\sum_{i\neq j}V|i><j|,                       \eqno(3)$$
    where $i$ labels the sites of a simple cubic $M\times M\times M$
    lattice. In the second sum only the nearest neighbor interactions
    are considered, and for sake of simplicity we chose $V=1$ as the
    unit of the energy scale. The potentials $\varepsilon_i$ are the
    site energies taken from a uniform distribution $-W/2\leq
    \varepsilon_i \leq W/2$. Therefore the disorder $W$ will be the
    critical parameter.

    One expects that for small disorder the spectrum of the
    Hamiltonian should be described by the level statistics of the GOE
    where due to hybridization level repulsion occurs and states
    become delocalized, while for large enough disorder the
eigenvalues
    will tend to be uncorrelated random numbers and the corresponding
    eigenstates will be localized. Therefore as disorder increases the
    MIT is accompanied with a transition from $P_{\rm GOE}(s)$ to
    $P_{\rm PE}(s)$ with some unknown spacing distribution $P_{\rm
    CE}(s)$ at the MIT. (The index CE stands for the critical
    ensemble occuring, as demonstrated below, at the MIT.) $P_{\rm
    CE}(s)$ may at the same time show characteristics of both the GOE
    and the PE as suggested by Shklowskii {\it et al.}$^1$ In infinite
    systems this transition is discontinuous;$^2$ however, simulations
    in finite systems show a continuous variation of the level spacing
    distribution. In fact there is a scaling property$^{1-3}$ of these
    $P(s)$ as $M$ changes for any fixed value of $W$. The sign of this
    scaling is clearly seen as a fixed ensemble in the $P(s)$, namely
    the CE obtained for different values of $M$. Moreover, there
    appears a fixed point $s_0\approx 2$ in the $P(s)$ curves for
    different disorders $W$. Therefore one may divide the interval
    $[0,\infty )$ into $[0,2]$ and $[2,\infty )$. The first part has
    been studied in Refs. 2, 3 and the latter (which is equivalent due
    to the normalization of $P(s)$) in Ref. 1. This time we will use
    all the numerically obtained $P(s)$ functions over a wide interval
    $s\in [0,5]$.

    The transition between the GOE and the PE can be approximated by
    several interpolation formulas. One of them is due to Brody$^6$
    $$P_{\rm B}(s)=c_1\,s^{\beta}\,\exp (-c_2s^{1+\beta}),  \eqno(4)$$
    where $c_1$ and $c_2$ are determined according to the conditions
    of normalization
    $$\int_0^{\infty}P(s)\,ds=1                             \eqno(5)$$
    and that the mean spacing is unity
    $$\int_0^{\infty}sP(s)\,ds=1.                           \eqno(6)$$
    Any spacing distribution $P(s)$ should satisfy Eqs. (5) and (6).
    Therefore we have $c_2=[\Gamma ((\beta +2)/(\beta +1))]^{1+\beta}$
    and $c_1=(1+\beta )\,c_2$. Another interpolation formula was given
    by Izrailev$^7$
    $$P_{\rm I}(s)= As^{\beta}(1+B\beta s)^Ce^{-Ds^2-Es},\eqno(7)$$
    where
    $$C={{2^{\beta}}\over {\beta}}\left (1-{\beta\over 2}\right
    )-0.16874, \qquad D={\pi^2\over 16}\beta,
    \qquad E={\pi\over 2}\left (1-{\beta\over 2}\right ),   \eqno(8)$$
    and the constants $A$ and $B$ are to be calculated numerically
    according to conditions (5) and (6). Both of these interpolations
    give back the limiting cases: for $\beta =1$ the GOE distribution
    and for $\beta =0$ the PE one.

    Concerning the $P(s)$ close to the MIT, in a recent publication
    Aronov {\it et al.}$^8$ have shown analytically that
    the distribution at the transition may well be described by
    $$P_{\rm A}(s)=c_1s\exp (-c_2s^{1+\beta}),              \eqno(9)$$
    where constants $c_1$ and $c_2$ are fixed according to conditions
    (5) and (6) and for parameter $\beta$ they obtained $0<\beta <1$.
    Furthermore $\beta$ is related to the correlation length exponent
    $\nu$ by$^9$
    $$\beta ={1\over {d\nu}}.                              \eqno(10)$$
    In this paper we wish to show a numerical analysis that confirms
    the form of Eq. (9) and at the same time provides a critical
    exponent that satisfies relation (10).

\bigskip
\centerline {\bf II. SHAPE ANALYSIS OF THE LEVEL SPACING DISTRIBUTION}
\medskip

    Since we expect to see a transition from the GOE to the PE
    statistics as disorder increases we propose to study such
    quantities that describe the shape of the calculated $P(s)$ and
    compare them to the known limiting cases. If one parameter scaling
    holds the plot of these quantities versus disorder obtained for
    different system sizes $M$ should show a fixed point, yielding the
    approximate position of the critical point as well as the
    approximate value of the critical exponent. For such a calculation
    Shklovskii {\it et al.}$^1$ used the tail, $s\in [2,\infty )$ of
    $P(s)$, while Hofstetter and Schreiber$^2$ employed the numerical
    fit of Eq. (7) with Eq. (8) on the other part, $s\in [0,2]$, of
    the $P(s)$. The latter authors have also analyzed$^3$ the
    integrated level statistics and the Dyson--Mehta statistics, as
    well, and shown that these quantities enable an even better finite
    size scaling then $P(s)$. In this contribution we introduce a
    different approach for the characterization of the level
    statistics. We will use all of our numerically obtained $P(s)$
    functions and in contrast to previous methods we will not
    introduce special parameters other than those that are uniquely
    related to the shape of the distribution function of a set of
    random numbers.

    It has already been shown$^{10}$ that it is advantageous to
    characterize a set of non--negative random numbers by certain
    moments of their distribution. This problem may arise studying
    e.g. noisy wave functions. The quantities introduced in Ref. 10
    are the spatial filling factor or participation ratio which is
    calculated as
    $$q={\mu_1^2\over\mu_2}                                \eqno(11)$$
    and the structural entropy
    $$S_{str}={\mu_S\over\mu_1}+\ln {\mu_2\over\mu_1},     \eqno(12)$$
    where $\mu_1$ and $\mu_2$ are the usual first and second moments
    of the distribution $p(x)$ of the random variables
    $$\mu_k=\int_0^{\infty }x^k\,p(x)dx,            \eqno(13{\rm a})$$
    and $\mu_S$ is calculated as
    $$\mu_S=-\int_0^{\infty }x\ln (x)\,p(x)dx.      \eqno(13{\rm b})$$
    As Eq. (6) ensures $\mu_1=1$ when using the level spacing
    distribution we will have simply $q=\mu_2^{-1}$ and $S_{str}=\mu_S
    + \ln \mu_2$. Note that in practical calculations Eqs. (13a) and
    (13b) are approximated by finite sums. The shape analysis resides
    on the comparison of points plotted on the $(q,S_{str})$ plane
    with curves calculated with known $p(x)$ functions.$^{11}$ Note
    that for a trivial distribution $p(x)=\delta (x-x_{0})$, one
    obtains $q=1$ and $S_{str}=0$. For any other distribution one will
    have $q\leq 1$ and $S_{str}\geq 0$ and the relations $0\leq q\leq
    1$ and $0\leq S_{str}\leq -\ln q$ always hold.

    The above characteristics will be employed here as well. Eqs. (11)
    and (12) should be calculated for every value of disorder
    parameter $W$ and system size $M$ replacing $x$ by $s$ and $p(x)$
    by $P(s)$. The calculated values $(q,S_{str})$ will be compared to
    the continuous curves obtained using the interpolation formulas
    due to Brody (4) and to Izrailev (7) as well as with other
    possible $P(s)$ functions. We will show that the $P_{\rm CE}(s)$
    is qualitatively different from the ones obtained for GOE and PE.

    We would like to emphasize that the calculation of $q$ and
    $S_{str}$ is a method which is not affected by the position of the
    fixed point in the $P(s)$ at $s\approx 2$. Therefore we are rather
    using all the obtained level distribution. At the same time no
    fitting procedure is necessary. We should note that one of the
    quantities, $q$, originally used as the participation ratio of a
    wave function serves as a measure of the skewness (or peakedness)
    of the distribution in our context. The structural entropy has no
    previously known meaning in our formalism. Additionally we have to
    mention that these quantities contain information about
    many--level correlations.$^4$

\bigskip
\centerline {\bf III. ANALYTICAL CALCULATIONS}
\medskip

    First we give the $(q,S_{str})$ relations for the interpolating
    distributions as we wish to compare the numerical results with
    these phenomenological functions. As $\beta$ runs from zero to
    unity the application of definitions (9) and (10) using $P_{\rm
    B}(s)$ from Eq. (4) yields explicitly
    $$q^{\rm B}(\beta)=\left [\Gamma\left ({{\beta +2}\over{\beta
    +1}}\right )\right ]^2\Bigg /\Gamma\left ({{\beta +3}\over{\beta
    +1}}\right ),                                          \eqno(14)$$
    and
    $$S^{\rm B}_{str}=\ln\Gamma\left ({{\beta +3}\over{\beta
    +1}}\right )-\ln\Gamma\left ({{\beta +2}\over{\beta +1}}\right )
    -{1\over{\beta +1}}\psi\left ({{\beta +2}\over{\beta +1}}\right ).
                                                           \eqno(15)$$
    Here $\psi (x)$ is the digamma function. On the other hand the
    $q^{\rm I}(\beta )$ and $S^{\rm I}_{str}(\beta )$ functions
    obtained using $P_{\rm I}(s)$ from Eqs. (7) and (8) can be
    calculated numerically with sufficient accuracy. Both of the cases
    are shown in Fig. 1, where the quantities, $q$ and $S_{str}$ are
    plotted as a function of $\beta$.

    As one can observe following the transition from PE to GOE the
    form of the $P(s)$ changes in two ways: first the low-$s$ behavior
    changes from a constant to linear level repulsion and at the same
    time the large-$s$ tail changes from $\exp(-s)$ to $\exp(-s^2)$.
    These two changes are accounted for by both of the interpolation
    functions (4) and (7). However, at the transition in our physical
    system one might expect that $P_{\rm CE}(s)$ shows characteristics
    of both of the two limiting ensembles. Keeping this in mind we
    introduce further possible interpolations between the exponential
    $P_{\rm PE}(s)$ and the Wigner $P_{\rm GOE}(s)$, e.g. such an
    intermediate distribution (IM1) may look like
    $$P_{\rm IM1}(s)=c_1\,s^{\beta}\exp (-c_2\,s),         \eqno(16)$$
    where $c_1=1+\beta$ and $c_2=c_1/\Gamma (1+\beta)$. The parameter
    $\beta$ runs in the $[0,1]$ interval. This distribution is that of
    the PE for $\beta =0$ and at $\beta =1$ it has the low-$s$
    behavior of the GOE. Similarly another intermediate distribution
    (IM2) may look like (see also Eq. (9))
    $$P_{\rm IM2}(s)=c_1\,s\exp (-c_2\,s^{1+\beta}),   \eqno(17)$$
    where $c_1=(1+\beta)[\Gamma (3/(1+\beta))]^2/[\Gamma
    (2/(1+\beta))]^3$ and $c_2=[\Gamma (3/(1+\beta))/\Gamma
    (2/(1+\beta))]^{1+\beta}$. In this formula the parameter $\beta$
    also runs in the $[0,1]$ interval. At $\beta =1$ this distribution
    is just the Wigner surmise and for $\beta =0$ it coincides with
    $P_{\rm IM1}(s)$ with $\beta =1$, i.e. the two functions meet with
    $$P_{\rm IM}(s)=4s\,e^{-2s},                           \eqno(18)$$
    which for $s\ll 1$ is a GOE and for $s\gg 1$ is a PE distribution.
    The quantities $q$ and $S_{str}$ for these new intermediate $P(s)$
    distribution as a function of their parameters are
    $$q^{\rm IM1}(\beta)={{\beta+1}\over{\beta+2}},        \eqno(19)$$
    $$S^{\rm IM1}_{str}(\beta)=\ln (\beta+2)-\psi (\beta+2)\eqno(20)$$
    for $P_{\rm IM1}(s)$ and
    $$q^{\rm IM2}(\beta)={{\left [\Gamma (3/(1+\beta))\right ]^2}\over
    {\Gamma (2/(1+\beta))\Gamma (4/(1+\beta))}},           \eqno(21)$$
    $$S^{\rm IM2}_{str}(\beta)=
              \ln\Gamma\left ({4\over{1+\beta}}\right )
             -\ln\Gamma\left ({3\over{1+\beta}}\right )
             -{1\over{1+\beta}}\psi\left ({3\over{1+\beta}}\right )
                                                           \eqno(22)$$
    for $P_{\rm IM2}(s)$. We have plotted Eqs. (19), (20) and Eqs.
    (21), (22) in Fig. 1. The combination of the two intermediate
    forms can be given as a two-parameter form
    $$P_{\rm IM3}(s)=c_1\,s^{\delta}\exp (-c_2\,s^{\alpha}),
                                                           \eqno(23)$$
    but for sake of simplicity we restrict ourselves to the
    one-parameter versions of either Eq. (16) or (17).

\bigskip
\centerline {\bf IV. NUMERICAL CALCULATIONS AND RESULTS}
\medskip

    In our investigation we have used the results of the numerical
    simulation presented and described in detail in Ref. 2. We have
    taken the data of the $P(s)$ histograms and calculated the
    quantities $q$ and $S_{str}$ as a function of $W$ around the
    critical disorder $15\leq W\leq 18$ for different system size $M$
    ranging from 13 up to 21. The states were obtained at the band
    center ($E=0$) for which the critical disorder is expected$^{12}$
    to be around $W_c=16.5$.

    First we present our results concerning the position of the
    critical point and the critical exponent. In Fig. 2 we show for
    $M=21$ how the calculated $q$ and $S_{str}$ values change with the
    increase of disorder interpolating between the PE and GOE values.
    For an infinite system one expects a step function--like behavior,
    here it is smeared out by the finite size of the system. In Fig.
    3 we have plotted our results for both quantities for different
    system sizes. The dashed line in both figures shows the expected
    position of the critical disorder. It is clear that a fixed point
    exists around $W_c=16.75$ for both quantities. In this work we
    have calculated the fixed point from second order polynomial fits
    to the data and averaged over the different pairs of $M$ and
    $M^{\prime}$. The value one obtains in this way is $W_c=16.87\pm
    0.52$ for $q$ and $W_c=16.77\pm 0.63$ for $S_{str}$.

    The critical exponent can be determined in a similar way. The
    approximate value is given by
    $$\nu_{M,M'}={{\ln (M/M')}\over{\ln (\Lambda_M/\Lambda_{M'})}},
                                                           \eqno(24)$$
    where
    $$\Lambda_M={{\partial X}\over{\partial W}}{\Bigg|}_{W_{M,M'}^c}.
                                                           \eqno(25)$$
    $W_{M,M'}^c$ is the approximate value for the critical
    disorder obtained for the pair of $M$ and $M^{\prime}$. The
    averaged results for the critical exponents are $\nu =1.27\pm
    0.29$ for $X=q$ and $\nu =1.30\pm 0.38$ for $X=S_{str}$.

    From these calculations we may conclude that this method does
    indeed give quantitatively correct results. The value of
    $\nu\approx 1.34$ for the critical exponent is obtained in Ref. 3.
    Note that the resulting value for the critical disorder $W_c$ is
    slightly higher than 16.5 as in recent calculations of the
    multifractal properties of the wave functions at the MIT.$^{13}$
    But noting the accuracy in both cases, we point out that
    $W_c=16.5$ is well within the error bars.

    Now we analyze the calculated $S_{str}$ values as a function of
    $q$ as $W$ changes around the critical point. Such a relation may
    be compared with the continuous ones obtained in the previous
    section. In Fig. 4 we display the data in the $S_{str}$ vs $q$
    diagram. The symbols denote the simulation and the curves the
    analytical results. The numerical data show a remarkable trend in
    that respect that they fall onto a common line independent of the
    system size for a wide range of disorder $15\leq W\leq18$. It is
    also clear that this trend is different from the $S_{str}(q)$
    curve observed for the Brody or the Izrailev distribution. In
    Fig. 5 we have enlarged the most important part of Fig. 4. As the
    numerical simulation leads to $(q,S_{str})$ values close to that
    of the approximation IM2 we conclude that the empirical
    $P(s)$ function should have very similar properties as the $P_{\rm
    IM2}(s)$ has. In fact choosing the calculated $P(s)$ function for
    $M=21$ and $W=16.75$, the corresponding quantities are $q\approx
    0.703$ and $S_{str}\approx 0.156$. These values can be obtained
    with the choice of $\beta =0.18\pm 0.02$ in Eq. (17), while for
    $W=16.5$ we compute $q\approx 0.708$ and $S_{str}\approx 0.153$
    which can be reproduced with $\beta =0.21\pm 0.02$ in Eq. (17).
    Hence we conclude that the intermediate distribution $P_{\rm
    IM2}(s)$ with a parameter $\beta\approx 0.20$ gives a good
    approximation of the $P_{\rm CE}(s)$ at the MIT. In Fig. 6 we show
    that the numerical histogram at $W=16.75$ for $M=21$ is well
    approximated by the distribution of the form of Eq. (17) with
    $\beta =0.2$.

    We note that this distribution shows the GOE characteristics for
    small level spacing $s$. For large $s$, however, does not follow
    the PE statistics (2) so that our data do not support the
    expectation that the CE shows characteristics of both limiting
    ensembles.

    The visible discrepancy between the curve for $P_{\rm IM2}$ and
    the calculated data in Figs. 4 and 5 is still unknown. We have
    performed preliminary calculations with the two--parameter
    distribution $P_{\rm IM3}$ given in Eq. (23). This function can
    approximate the numerical points with a better accuracy, e.g. with
    the choice $\delta\approx 1.3$ and $\alpha\approx 1.1$. Such
    situation with $\delta >1$, however, would violate a general
    symmetry theorem by Dyson.$^{14}$ We have also performed a very
    accurate analysis of small--$s$ data considering the integrated
    level spacing distribution and obtained a value $\delta\approx
    0.97$ for the best fit at the MIT.$^{15}$

    The result presented is at the same time capable to explain the
    strange behavior of the normalization parameter $A$ (see e.g. Eq.
    (7)) observed in Ref. 2, where at the transition Hofstetter and
    Schreiber reported indications of a discontinuous change of $A$ as
    a function of $W$. The normalization constant in Eq. (17) with
    $\beta=0.2$ is $c_1\approx 2.28$ which is larger than for the
    Poisson and Wigner distributions. In the limit $M\to\infty$ the
    normalization constant is expected to be $A=\pi/2$ on the metallic
    side and $A=1$ on the insulating side while at the transition it
    is larger than both values. Similar arguments for the parameter
    $B$ in Eq. (7) follow from Eqs. (7) and (8) and the above
    arguments for $A$, explaining the respective observations in Ref.
    2.

\bigskip
\centerline {\bf V. CONCLUSIONS}
\medskip

    We have presented a general method for the analysis of the spacing
    distribution around the critical point of a second order phase
    transition. As an example we have calculated the position of the
    MIT and the critical exponent in the 3D Anderson model. Although
    with lower accuracy, this method does indeed give the correct
    answer. On the other hand the main result of our paper is that for
    the MIT in the 3D Anderson model we have found a possible shape of
    the spacing distribution $P_{\rm CE}(s)$ as given by Eq. (17).
    We have also presented an explanation for the strange behavior of
    the normalization constant observed in Ref. 2. Our results are in
    agreement with recent theoretical expectations derived by Aronov
    {\it et al.}$^8$ which yield$^9$ the relation (10) between $\beta$
    and the correlation length exponent $\nu$. For our numerical value
    $\nu\approx 1.3$ in $d=3$ the relation (10) yields a value
    $\beta=0.26$, which is close to the value $\beta\approx 0.2$,
    obtained from the shape analysis above.

\bigskip
\centerline {\bf ACKNOWLEDGEMENT}
\medskip
    One of the authors (I.V.) is grateful for the warm hospitality at
    the Institut f\"ur Physikalische Chemie,
    Johannes-Gutenberg-Universit\"at where part of this work has been
    completed. Financial support from Orsz\'agos Tudom\'anyos
    Kutat\'asi Alap (OTKA), Grant Nos. 517/1991, T7238/1993 and
    T014413/1994 is gratefully acknowledged, as well as from the
    Deutscher Akademischer Austauschdienst.
\vfill\eject
\baselineskip=.8truecm plus .1truecm
    {\bf References}
\medskip
\leftskip 1truecm

\item{$^*$}Present address: Blackett Laboratory, Imperial College,
    London SW7 2BZ, UK

\item{$^1$}D. I. Shklovskii, B. Shapiro, B. R. Sears, P. Lambrianides,
    and H. B. Shore, Phys. Rev. B {\bf 47}, 11487 (1993).

\item{$^2$}E. Hofstetter and M. Schreiber, Phys. Rev. B {\bf 48},
    16979 (1993).

\item{$^3$}E. Hofstetter and M. Schreiber, Phys. Rev. B {\bf 49}
    (1994) in press.

\item{$^4$}M. L. Mehta, {\it Random Matrices} (Academic Press,
    Boston, 1991).

\item{$^5$}D. R. Grempel, R. E. Prange, S. Fishman, Phys. Rev. A {\bf
    29}, 1639 (1984).

\item{$^6$}T. A. Brody, Lett. Nuovo Cimento {\bf 7}, 482 (1973).

\item{$^7$}G. Casati, F. Izrailev, and L. Molinari, J. Phys. A: Math.
    Gen. {\bf 24}, 4755 (1991).

\item{$^8$}A. G. Aronov, V. E. Kravtsov, I. V. Lerner, JETP Lett. {\bf
    59}, 39 (1994).

\item{$^9$}V. E. Kravtsov, I. V. Lerner, B. L. Altshuler, and A.
    G. Aronov, Phys. Rev. Lett. {\bf 72}, 888 (1994).

\item{$^{10}$}J. Pipek and I. Varga, Phys. Rev. A {\bf 46} 3148,
    (1992).

\item{$^{11}$}I. Varga, Ph. D. Thesis, Technical University of
    Budapest (1993), unpublished.

\item{$^{12}$}B. Bulka, K. Broderix, A. MacKinnon, and M. Schreiber,
    Physica A {\bf 167}, 163 (1990).

\item{$^{13}$}H. Grussbach, private communication.

\item{$^{14}$}F. J. Dyson, J. Math. Phys. {\bf 3}, 140 (1962).

\item{$^{15}$}E. Hofstetter and M. Schreiber, unpublished.

\vfill \eject
    {\bf Figure Captions}
\medskip
\item{\bf Fig. 1.} Quantities $q$ and $S_{str}$ as functions of the
    free parameter $\beta$ for the Brody (dashed line) the
    Izrailev distribution (solid line), the first intermediate
    (IM1) (dashed--dotted line), and the second intermediate (IM2)
    distribution (dotted line).

\item{\bf Fig. 2.} Calculated $q$ and $S_{str}$ for $M=21$ as a
    function of disorder $W$. The limiting values for the GOE and PE
    are shown with horizontal dashed--dotted lines. The expected
    position of the MIT is shown by a vertical dashed line.

\item{\bf Fig. 3.} Calculated $S_{str}$ (a) and $q$ (b) for
    $M=13,15,17,19,21$ as a function of disorder $W$. The expected
    position of the MIT is shown by a vertical dashed line.

\item{\bf Fig. 4.} Calculated $S_{str}$ vs $q$ for $M=13,15,17,19,21$
    in the range of $15\leq W\leq 18$. The solid circles are the
    points representing the PE, the GOE and the IM (Eq. (18)) cases.
    Solid line represents the relation for the Izrailev, dashed line
    the one for the Brody distribution. The dashed--dotted line
    reflects the intermediate distribution IM1 (between PE and IM)
    given in Eqs. (19), (20), and IM2 (between IM and GOE) given in
    Eqs. (21), (22).

\item{\bf Fig. 5.} A part of Fig. 4 enlarged. The result of the
    simulation at $W=16.75$ for $M=21$ is plotted bold and marked with
    the text MIT. The solid line is obtained from the interpolation
    with the Izrailev, dashed line with the Brody, and dashed--dotted
    line with the intermediate (IM2) distribution.

\item{\bf Fig. 6.} The numerically obtained histogram of $P(s)$ for
    $W=16.75$ and $M=21$ (solid line) and the distribution given in
    Eq. (17) with $\beta =0.2$ (dashed line).
\bye